# Perceptions of Gender Diversity's impact on mood in software development teams

**Kelly Blincoe,** Department of Electrical, Computer, and Software Engineering, University of Auckland
**Olga Springer,** Faculty of Electronics, Telecommunications and Informatics, Gdansk University of Technology
**Michał R. Wróbel,** Faculty of Electronics, Telecommunications and Informatics, Gdansk University of Technology

**Abstract**: Recent studies show that gender diversity in IT teams has a positive impact on the software development process. However, there is still a great gender inequality. The aim of our study was to examine how the working atmosphere depends on the gender differentiation of IT teams. The analysis of the results of the interviews and questionnaires showed that the atmosphere in gender-differentiated teams is more pleasant compared to purely male ones. The paper also discusses the problem of gender discrimination, which, according to the results of the study, unfortunately still exists and affects the working atmosphere. Finally, we looked at ways to reduce the gender inequity, where it turned out that soft approaches such as dedicated training, workshops to show the human face of the IT industry are preferred.

## 1. Introduction

Research conducted in recent years has shown that emotions and moods can have a serious impact on the work of software development teams [1]. To shed new light on the issues affecting the working atmosphere, we conducted a study to examine perceptions whether and how gender diversity impacts on the mood and atmosphere, and thus on the comfort of IT teams.

We also explored the problem of gender discrimination, which, if it occurs, has a significant impact on the mood, emotions of employees, and the working atmosphere in general.

The study was motivated by the fact that gender inequality still exists in the IT sector [2]. According to the National Center for Women & Information Technology (NCWIT), women in IT in the USA constitute only 26% of all employees [3]. In the EU, this percentage is even lower and reaches only 17.2% of ICT sector employees [4]. Many companies, especially large ones, are aware of the problem and are actively trying to reduce the gender gap by conducting activities and campaigns to encourage women to work in the software industry and to create a better workplace free from prejudice and discrimination. Examples of such actions include Microsoft's Global Diversity and Inclusion [1], Google's Diversity[2] or Mozilla Diversity and Inclusion Strategy[3].

Bottom-up initiatives are also emerging. They are mostly initiated by women who already work in the software industry. Their main goal is to bring together women interested in information technology to support each other. An example of such a social activity is a Facebook group called "Women in Tech", which is followed by more than 1 million of users. Numerous events are also organized at universities to encourage girls to choose a technical career path. Among other things, the goal of these events is to show that the IT sector can be a friendly working environment. However, recent studies have suggested that the ICT sector can be unfriendly towards women [5]. To increase gender diversity, this must be changed. In Software Engineering, gender diversity has been shown to increase innovation [6], increase productivity [7][8], reduce turnover [7], reduce conflict within teams [9], and produce more user-friendly software [10]. Thus, the culture of the IT workforce, related to gender diversity, must be studied more deeply to better understand the issues surrounding gender diversity, in the hopes to improve it. Our findings show that gender diversity is associated with a more pleasant team mood, but barriers still exist in achieving gender diversity.

---

1 https://www.microsoft.com/en-us/diversity
2 https://diversity.google/
3 https://wiki.mozilla.org/Diversity_and_Inclusion_Strategy

The study was carried out in two stages. In the first stage 16 interviews with members of IT teams were conducted. They all worked in software companies located in the Pomeranian region of Poland. Then, on the basis of the collected data, questions for the survey were developed. For those interested in replicating this study, the full questionnaire is available at [11].

The questionnaire study was conducted for 3 weeks and 252 participants took part in it. Although the survey was addressed to an international audience, the vast majority (90.9%) of participants work in Poland. The participants were nearly equally split between women and men with responses from 125 women, 123 men, and 4 people who described their gender as other. The results of the survey are summarized in Figure 1 and discussed in detail in the remaining sections.

**Limitations of the study:** The survey captures perceptions of practitioners. Since the participants were aware that the focus of the survey was on gender differences, it is possible this influenced their responses and did not accurately capture their perceptions [12]. However, we did find interesting differences between the male and female responses. Another threat to external validation is that the majority of respondents work in Poland. We do not claim generalizability and plan to expand this survey to a wider audience in future research.

## 2. Workplace atmosphere

**Acceptance of Women:** The vast majority of respondents (77.4%) positively assess the trend of increasing the number of women in IT. While some are neutral on this topic, only 4.8% of respondents disagreed. **Gender Diversity Impact on Atmosphere:** During the interviews, male participants commonly stated that the working atmosphere in diverse teams is more pleasant. On the other hand, it was often stated that in purely male teams the atmosphere is more harsh with more indecent jokes and rude cut-outs. Participants believed such behaviour was reduced when gender diversity was increased. These observations were also confirmed by the survey. Nearly 59.8% of men confirmed that they experienced such behaviour while working on male-only teams. Such answers suggest that a more friendly atmosphere in work environment can be found in gender diverse teams. During the interviews it was suggested that with women on board, there is less aggression, a milder atmosphere, and better hygiene at work.

Only 12.3% of men reported they would prefer to work in purely male teams indicating that perhaps men do not like the harsh atmosphere it brings. Future research should investigate this more directly.

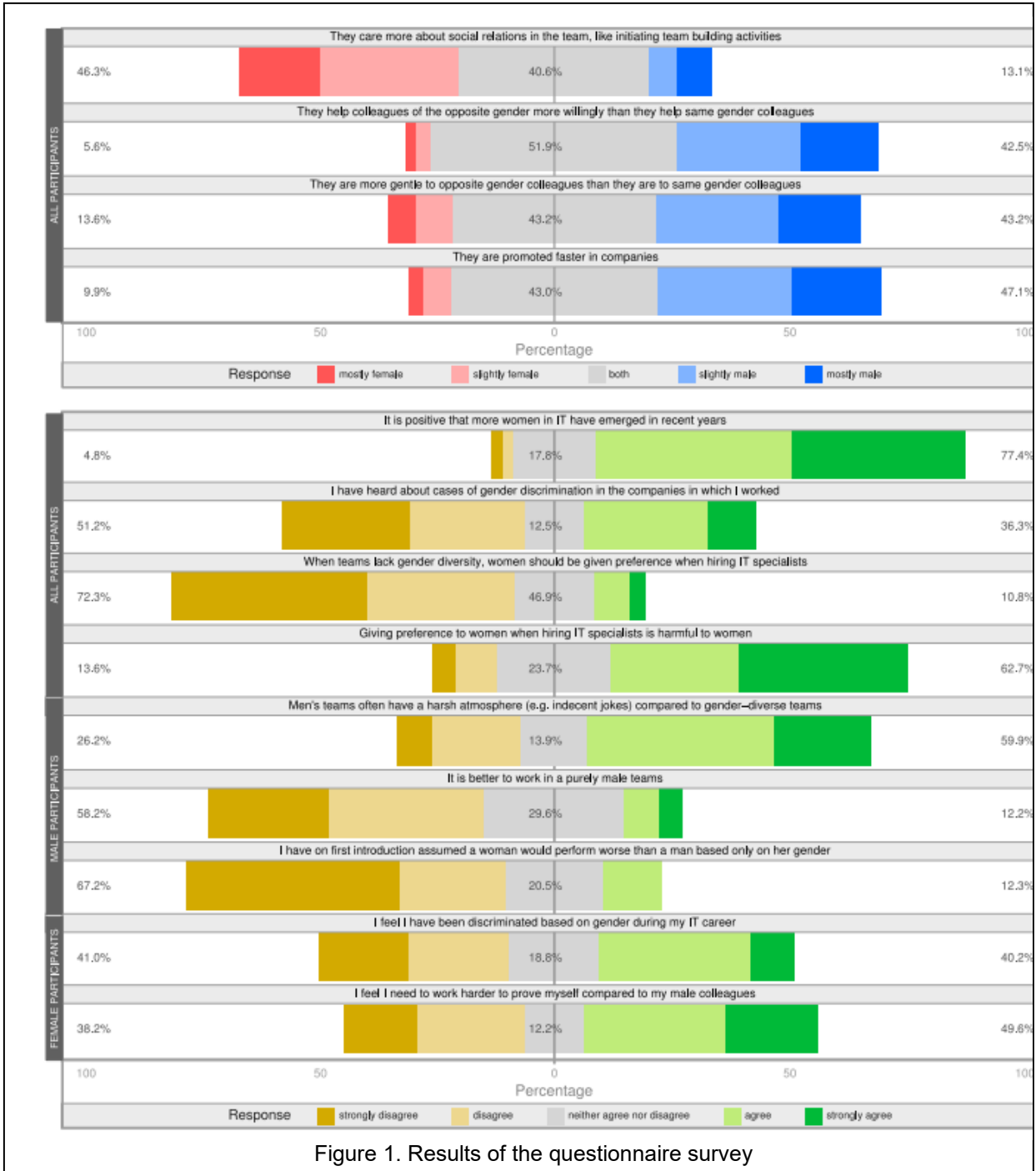

Figure 1. Results of the questionnaire survey

**Social Relationships:** Women are perceived by study participants as being more active in building social bonds among employees. During the interviews, some participants claimed that women have a greater passion for team integration, such as initiating and organizing meetings after work. There have even been suggestions that women have a greater need to take care of team morale. Again, this was confirmed by a survey where both male and female respondents reported women were better at creating cohesion within the teams. Only 13% of respondents indicated men as more active in the broadly understood social relations, while women were indicated by as much as 46.5%. Others

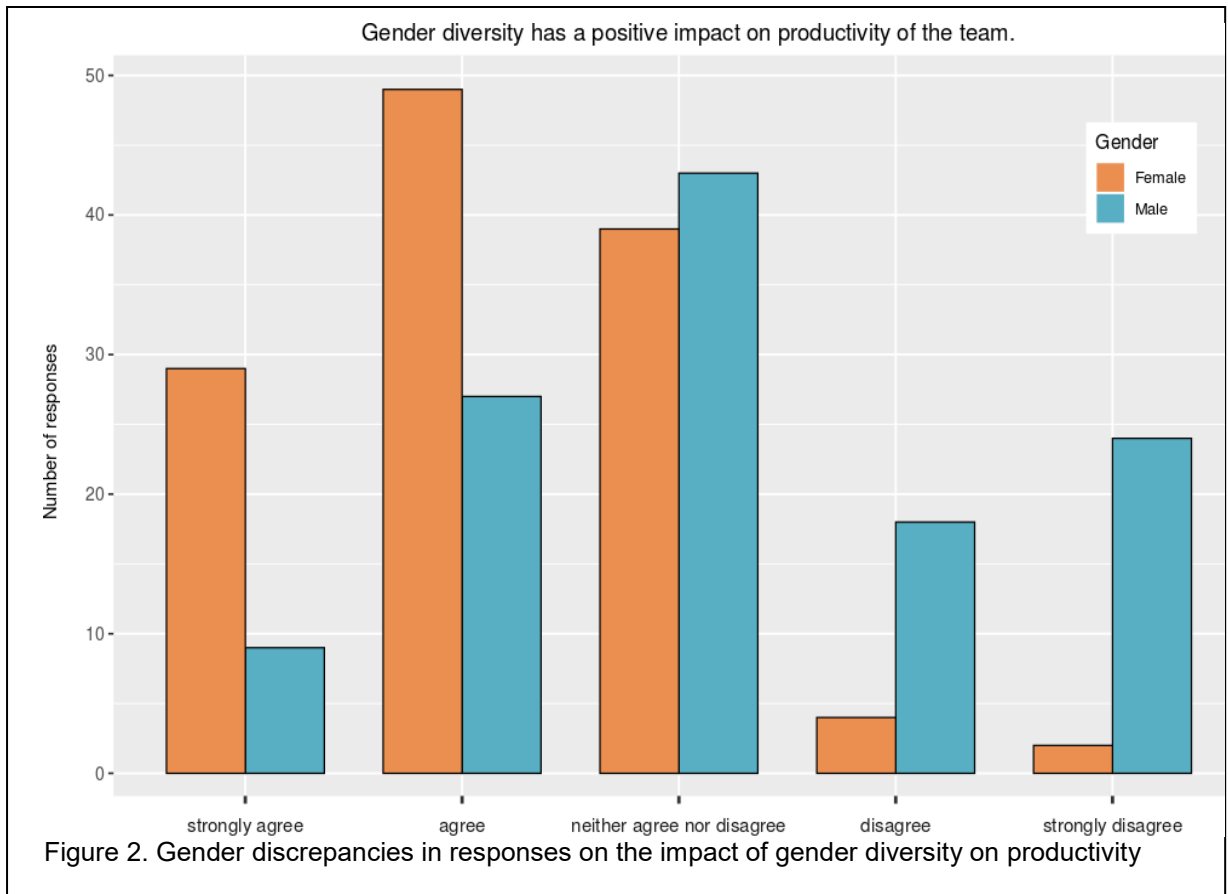
Figure 2. Gender discrepancies in responses on the impact of gender diversity on productivity

considered these issues to be gender independent. While this indicates that increasing gender diversity can ensure stronger team cohesion, managers must be careful not to place the responsibility for creating team bonds on female team members, who may not want to fill this role. Past research has shown that women are often expected to fill these nurturing roles, while men are expected to be ambitious [13].

The working climate is undoubtedly also influenced by the way in which help is provided to co-workers. During the interviews, the participants repeatedly suggested that colleagues are much more willing to help women than men. If a man turns to another man for help, he is more often treated with impatience and may even be rejected. Such behaviour often spoils the atmosphere in the whole team. The survey confirmed that it is easier for women to get help. In addition, 50% of men confirmed that they are more willing to help women, while for the rest, gender is irrelevant. Furthermore, men are much more often (43.2%) perceived as more gentle to females than to other males. Again, this could point to gender stereotypes in IT teams where men are perceived as more competent, and, thus, do not need help [14].

**Gender Diversity Impact on Work:** We also asked respondents of the survey whether gender diversity in IT teams has a positive impact on work - whether it increases innovation, productivity and product quality. While past work has proven that gender diversity does in fact have positive impacts on each of these, the perceptions of the respondents are mixed. The results showed large gender discrepancies in the results

(Fig. 2). Men's answers were almost evenly distributed: 35.5% did not have an opinion on this issue, 34.7% definitely do not see such an improvement, while 29.7% see an improvement of the work caused by gender diversity. On the other hand, most women (over 60%) do believe gender diversity brings positive benefits. This is likely something that can be improved.

## 3. Gender discrimination

There are many well-known cases of gender discrimination in the IT community[4]. As cases of discrimination affect both the mood of employees and the overall work atmosphere, part of the study was devoted to this problem.

**Amount of Discrimination:** None of the women interviewed confirmed that they had been discriminated against, and none of the participants, both women and men, had encountered this problem during their careers. On the other hand, the situation among the survey participants is diametrically different. As many as 40.2% of women agreed with the statement "I feel I have been discriminated based on gender during my IT career". These results indicate that cases of discrimination are likely underreported when women are required to report these events in a face-to-face setting (like our interviews).

**Awareness of Discrimination:** There are some interesting discrepancies around awareness of discrimination. Only 25.4% of men report to have heard about discrimination in their work environment, compared to 52.4% of women. This could be because women are less likely to share stories of discrimination with their male co-workers.

**Conscious Biases:** In our survey, only 12.3% of male respondents admit to assuming a female coworker would perform worse than a male counterpart upon first introduction. Yet, during the interviews, women reported that upon joining a new team as the first female member, they felt the male team members were unsure of how they would manage on the team. In addition, nearly half of women (49.6%) feel they have to work harder to prove their worth compared to their colleagues. This indicates that there could be either some unreported subconscious biases among men (often referred to as the "Prove it again" bias [15]) or that women perceive such a bias when one doesn't exist (for the majority of male respondents). Even so, 12.3% of male's admitting a conscious bias is still problematic and should be addressed.

**Opportunity for Advancement:** Almost half (47.3%) of respondents perceive that men are promoted faster than women. However, among men's responses, this percentage is only 25.6%. Most men (61.5%) think that gender does not affect promotion, while 68.3% of women feel gender does affect promotion in favour of men.

**Favoritism:** Many respondents (44.2%) do not believe there are differences related to gender and management preferences. However, of those who do report they feel gender influences the opinions of their management, both women and men report that they believe the opposite gender is more favored by management, indicating there are clearly some skewed perceptions around favoritism.

## 4. Reducing the gender gap

Many companies are currently conducting various campaigns to reduce gender gaps. They take different forms, from providing training for women or scholarships dedicated to female students, to strictly defined employment quotas.

---

4 https://geekfeminism.wikia.org/wiki/Timeline_of_incidents

**Hiring Women to Increase Gender Diversity:** During the interviews, both women and men were strongly critical of gender preferences in employment. Women in particular claimed that such situations in the company have a negative impact on their image as specialists. Some even claimed that they feel they are perceived as those who got a job, not because of their professional qualifications, but mainly because they are women. This was also confirmed by the survey results. More than half of women (59.6%) are against gender preference in the employment process, and only 16.9% support such mechanisms. Also, the majority (62.6%) of respondents agree that such preferences when hiring IT specialists is detrimental to women.

**Reducing Stereotypes:** As a way to reduce the gender gap, the women we interviewed mainly pointed to the need to change the image of software engineers commonly established as the male world of IT geeks. In addition, they mentioned the need for changes already at primary school level, including better teaching of mathematics as well as raising girls' parents awareness of the advantages of technical education. One interlocutor mentioned that only her brother in childhood was encouraged by the parents to learn programming, while both showed the same technical skills.

# 5. Conclusions

Our study showed that the majority of IT sector employees perceive the atmosphere in gender-diverse teams to be more pleasant. Only a few men reported they would prefer to work in a purely male environment. However, there is still much to be done to ensure equality. A significant number of women have experienced gender discrimination during their careers and more than 12% of men reported having conscious biases against women. Further research is needed on how these issues can be resolved.